\documentclass{Interspeech}

\interspeechcameraready

\title{Towards Emotionally Consistent Text-Based Speech Editing: \\Introducing EmoCorrector and The ECD-TSE Dataset}

\author[affiliation={1}]{Rui}{Liu}
\author[affiliation={1}]{Pu}{Gao}
\author[affiliation={1}]{Jiatian}{Xi}
\author[affiliation={2}]{Berrak}{Sisman}
\author[affiliation={3}]{Carlos}{Busso}
\author[affiliation={4,5}]{Haizhou}{Li}

\affiliation{Inner Mongolia University}{Hohhot}{China}
\affiliation{Center for Language and Speech Processing}{Johns Hopkins University}{USA}
\affiliation{LTl}{Carnegie Mellon University}{USA}
\affiliation{SRIBD, School of Data Science}{The Chinese University of Hong Kong}{Shenzhen, China}
\affiliation{Department of ECE}{National University of Singapore}{Singapore}

\email{liurui\_imu@163.com}


\usepackage{comment}
\usepackage{graphicx}
\usepackage{subcaption}
\usepackage{multirow}
\usepackage{float}  
\usepackage{tabularx}
\usepackage{caption}
\usepackage{ulem}
\usepackage{svg} 
\usepackage[most]{tcolorbox} 

\makeatletter
\renewcommand{\anonname}{%
    \ifinterspeechfinal
        \authorlist
        \vspace{-2.7mm}  
        \par
        \hypersetup{
            pdftitle={\@title},
            pdfauthor={\metadataauthors},
            pdfkeywords={\@keywords},
            pdfinfo={Affiliation=\metadataaffiliations}
        }
    \else
         Anonymous submission to Interspeech 2025
    \fi
}
\renewcommand{\anonaddress}{%
  \ifinterspeechfinal
    \begingroup
      \parskip=-0.1em 
      \affiliationslist
    \par\endgroup
  \fi
}
\makeatother

\begin{document}

\maketitle

\begin{abstract}
Text-based speech editing (TSE) modifies speech using only text, eliminating re-recording. However, existing TSE methods, mainly focus on the content accuracy and acoustic consistency of synthetic speech segments, and often overlook the emotional shifts or inconsistency issues introduced by text changes. To address this issue, we propose EmoCorrector, a novel post-correction scheme for TSE. EmoCorrector leverages Retrieval-Augmented Generation (RAG) by extracting the edited text's emotional features, retrieving speech samples with matching emotions, and synthesizing speech that aligns with the desired emotion while preserving the speaker's identity and quality. To support the training and evaluation of emotional consistency modeling in TSE, we pioneer the benchmarking Emotion Correction Dataset for TSE (ECD-TSE). The prominent aspect of ECD-TSE is its inclusion of $<$text, speech$>$ paired data featuring diverse text variations and a range of emotional expressions. Subjective and objective experiments and comprehensive analysis on ECD-TSE confirm that EmoCorrector significantly enhances the expression of intended emotion while addressing emotion inconsistency limitations in current TSE methods. \textcolor[rgb]{0.93,0.0,0.47}{Code and audio examples are available at \url{https://github.com/AI-S2-Lab/EmoCorrector}}.
\end{abstract}

\section{Introduction}
Text-based Speech Editing (TSE)
modifies audio by editing its underlying text rather than the audio signal directly. With the rise of digital media, TSE has become essential for applications like social media content creation
, game voiceovers
, and film dubbing
, as it corrects issues such as mispronunciations, omissions, or stuttering without requiring a full re-recording \cite{jiang2023fluentspeech}.



\begin{figure}[t]
    \centering
    \includegraphics[width=0.8\linewidth]{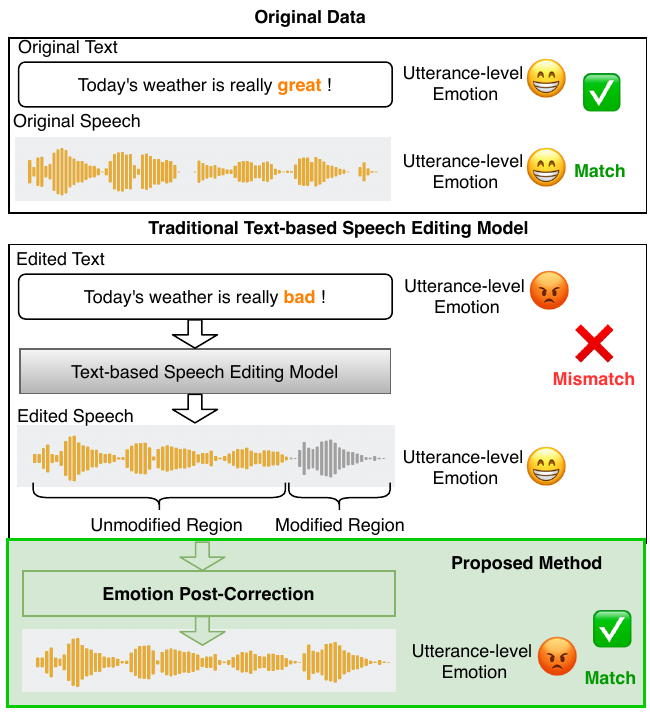}
    \caption{Our approach lies in correcting the emotional mismatch or inconsistency issue of traditional TSE methods.}
    \label{fig1}
    \vspace{-2em}
\end{figure}

Recent advancements in text-to-speech (TTS)
have led to the development of neural models for TSE \cite{jiang2023fluentspeech,bai20223,liu2023fluenteditor,wang2024speechx,DBLP:conf/acl/Peng00MH24}. For example,
Despite these advancements, most TSE methods focus on content modification and acoustic quality, often neglecting emotional consistency \cite{wei2023tackling}.
As shown in Fig. \ref{fig1}, although there is only one word difference between the edited text and the original text, the emotional expression at the sentence level is very different. Although the traditional TSE model successfully synthesizes the speech segment for the word ``bad", it does not express the emotion that the edited text should contain. At this point, the emotional state of the whole sentence needs to be further corrected to achieve emotional consistency.

A natural idea is to use \textit{Emotional Voice Conversion (EVC) }\cite{zhou2022emotional,9829916} models to transform the emotion of an entire sentence into the same emotional state that the edited text contains without altering the speaker's identity.
However, directly using an EVC model presents several issues. 1) Before applying EVC, it is necessary to identify the emotional state embedded in the edited text, a process that may introduce errors. 2) Even if the identified emotion state is accurate, using discrete emotion labels as input for the EVC model may fail to capture rich and nuanced emotional information, thereby limiting the model's performance. 3) The output of the EVC model is inherently uncontrollable, making it difficult to evaluate whether the converted results truly achieve optimal performance. Therefore, directly relying on an EVC model for post-correction does not fully meet our requirements.

To address the above issues, we propose an end-to-end post-correction approach to rectify emotional inconsistencies for TSE, termed EmoCorrector.
Our EmoCorrector is inspired by the concept of Retrieval-Augmented Generation (RAG) \cite{lewis2020retrieval}. First, emotional embedding are extracted from the edited text, and speech samples with similar emotional embedding are retrieved from the cross-modal retrieval database. Next, the emotional embeddings of these retrieved speech samples are used as a reference, while the speaker features from the original edited speech are retained to preserve speaker identity. Finally, TTS synthesis is performed, enabling the modification of the original edited speech's emotional information and achieving consistent emotional expression.
For the feature construction of the cross-modal database, we adopt a contrastive learning \cite{meng2023calm} strategy focused on emotional features to perform language-audio contrastive learning, aligning the emotional features of text and speech within a unified space. Additionally, we design a speaker-emotion feature disentanglement strategy \cite{li2022cross} based on adversarial learning \cite{hsu2019disentangling} to ensure the independence of emotional and speaker feature representation.

To support training and evaluation for emotional post-correction, we construct a new Emotion Correction Dataset for TSE (ECD-TSE). This dataset includes original text as well as edited text reflecting different emotional states. We utilized three advanced TTS models to perform emotional speech synthesis for all the text. This emotion-rich speech database, containing text editing information, provides a solid foundation for modeling emotional consistency.
Extensive experiments on ECD-TSE demonstrate that EmoCorrector significantly enhances emotional alignment while maintaining natural audio quality.
The main contributions of this paper can be summarized as follows:
1) We propose a comprehensive end-to-end emotion post-correction scheme for TSE, named EmoCorrector. 2) Our EmoCorrector draws on the techniques of RAG and designs cross-modal emotion contrastive learning and speaker-emotion disentanglement learning strategies. 3) We introduce the groundbreaking ECD-TSE dataset, which advances emotional consistency modeling and evaluation in the field of TSE.

\vspace{-0.7em}
\section{Dataset: ECD-TSE}

For the emotionally consistent modeling of TSE, we need a database where sentences conveying similar lexical information with few modifications elicit different emotions. Unfortunately, popular databases such as the ESD dataset \cite{zhou2021seen}, MSP-PODCAST \cite{duret2024msp}, and IEMOCAP \cite{busso2008iemocap} are not appropriate for this task since they are usually a fixed text corresponding to the speech with various emotions.
To this end, we develop the ECD-TSE corpus through two crucial steps: \textit{Text variant generation} and \textit{Emotional speech generation}.

\textbf{Text Variant Generation:} The process of text variant generation is to generate neutral text first, and then generate text variants with different emotions.

First, we instruct ChatGPT-4 to generate texts with a neutral emotion. Our prompt for this step is as follows: \uline{``Please help me to build a text dataset, with each sentence consisting of 4-10 words, and in English. Please write 10 sentences, and ensure the text contains words that express neutral emotions."}
Subsequently, we instruct ChatGPT-4 to modify the neutral texts by altering one or two words, thereby inducing a change in the text's emotional connotation. Our prompt for this modification is as follows (The following example takes ``sadness'' as an example): \uline{``Now, I will give you a sentence. Please modify only one or two words to change the emotion to {sadness}. Please output only one modified sentence."}

As shown in Table~\ref{tab:emotion_text}, in this manner, we obtain texts corresponding to four additional emotional states, with only one or two words differing between the various emotions.

\begin{table}[t]
\centering
\vspace{-0.8em}
\caption{Examples of text data. There are only one or two words different between different texts, but their emotional expressions are completely different.}
\vspace{-0.8em}
\begin{tabular}{c|l}
\toprule
\textbf{Emotion} & \multicolumn{1}{c}{\textbf{Text}} \\
\hline
Neutral & My sister studies in the library. \\
Sadness & My sister \textbf{cries} in the library. \\
Angry & My sister \textbf{yells} in the library.  \\
Happy & My sister studies in the \textbf{joyful} library.  \\
Fear & My sister studies in the \textbf{haunted} library.  \\
\bottomrule
\end{tabular}
\vspace{-1em}
\label{tab:emotion_text}
\end{table}

\textbf{Emotional Speech Generation:} For emotional speech generation, we utilize three advanced expressive TTS systems, that are Microsoft Azure \cite{azure_speechstudio}, CosyVoice2 \cite{du2024cosyvoice}, and F5-TTS \cite{chen2024f5}, as the synthesizer. For Microsoft Azure, when we synthesize each sentence of text, we directly set the emotion label corresponding to the text in the engine, randomly set the speaker identity, and it can synthesize the highly expressive speech of a specific speaker corresponding to the emotion.
For Cosyvoice2 and F5-TTS, a reference is required for emotional speech synthesis. Therefore, we randomly select speech samples with matching emotional labels from third-party emotional datasets as references based on the emotional labels of the text to be synthesized. Note that the speaker identity of the synthesized speech is consistent with the reference speech. For the third-party emotional dataset, we use the ESD \cite{zhou2021seen} and MEAD \cite{wang2020mead} datasets, both of which provide diverse emotional speech samples for diverse speaker identities.
Ultimately, the Azure system synthesizes speech for 5 speakers, CosyVoice2 synthesizes speech for another 5 speakers, and F5-TTS synthesizes speech for an additional 2 speakers. The statistic information is shown in Table~\ref{tab:dataset_info}.



\begin{table}[t]
\centering
\vspace{-0.8em}
\caption{Statistics information of the ECD-TSE.}
\vspace{-1em}
\begin{tabular}{c|c}
\toprule
\textbf{Attribute} & \textbf{Value} \\
\hline
Speech Samples & 84,000 = 1,400 * 5 * 12 \\
Emotions & 5 (Happy, Sadness, Neutral, Angry, Fear) \\
Speakers & 12 (6 male and 6 female) \\
Text & 7,000 = 1,400 * 5 \\
Total Duration & Approximately 90 hours \\
Sampling Rate & 16,000 Hz \\
\bottomrule
\end{tabular}
\vspace{-2.2em}
\label{tab:dataset_info}
\end{table}

\begin{figure*}[t]
    \centering
    \includegraphics[width=0.8\textwidth]{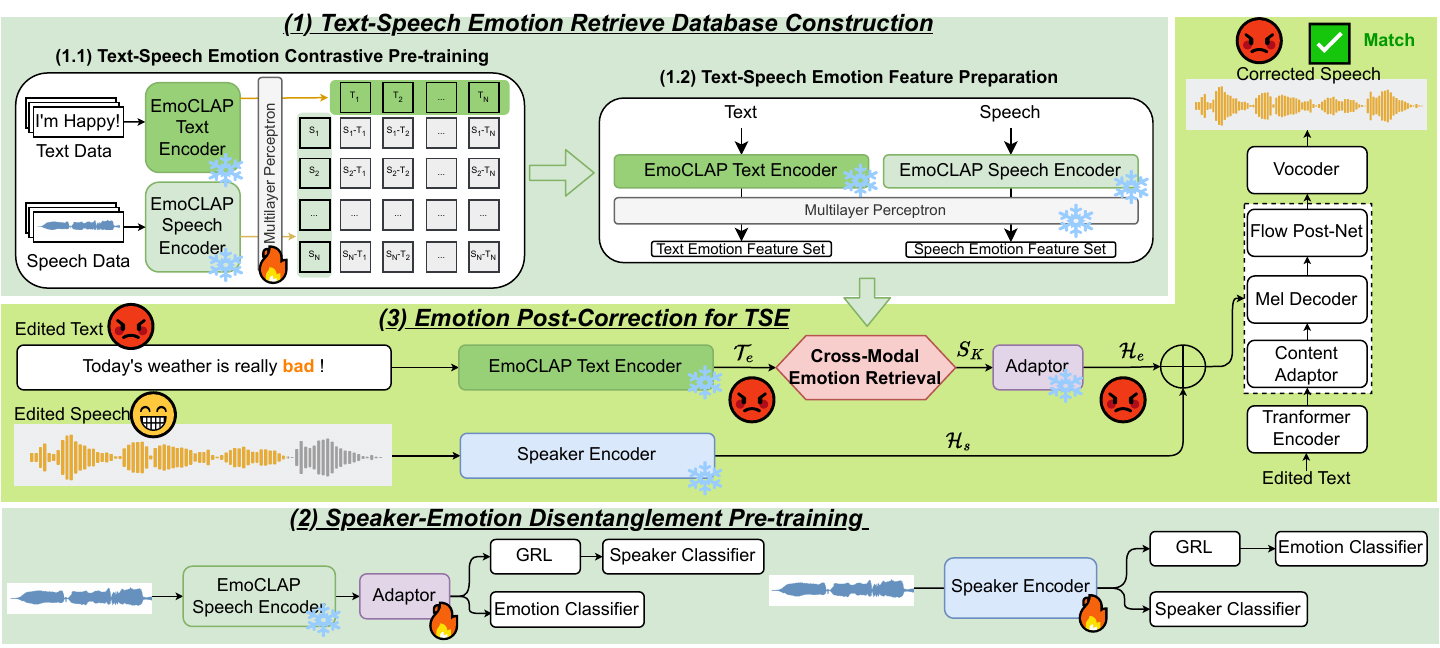}
    \vspace{-0.8em}
    \caption{The overall workflow of EmoCorrector.
    }
    \label{workflow}
    \vspace{-2em}
\end{figure*}

\vspace{-0.8em}
\section{Methodology: EmoCorrector}

Our proposed {EmoCorrector} framework ensures emotional consistency in TSE while preserving speaker identity through three stages:  1) Text-Speech Emotion Retrieval Database Construction (Block 1 in Fig.2), 2) Speaker-Emotion Disentanglement Pre-training (Block 2 in Fig.2), and 3) Emotion Post-Correction for TSE (Block 3 in Fig.2). The first two modules require pre-training, and then the third module is trained end-to-end.


\subsection{Text-Speech Emotion Retrieve Database Construction}
\textbf{Text-Speech Emotion Contrastive Pre-training: }We propose Emotional Contrastive Language-Audio Pretraining (EmoCLAP), inspired by CLAP~\cite{elizalde2023clap}, to learn a shared semantic space for text and speech emotion representations. As shown in Fig.~\ref{workflow} (Block 1.1), the EmoCLAP Text Encoder (based on pre-trained RoBERTa~\cite{liu2019roberta}) extracts text emotional features $T$, while the Speech Encoder (built on emotion2vec~\cite{DBLP:conf/acl/MaZYLGZ024}) extracts emotional speech features $S$. Both are projected into a common embedding space via a multi-layer perceptron (MLP):
$\mathcal{M}_s = \text{MLP}(S), \quad \mathcal{M}_t = \text{MLP}(T)$.

To enhance text-speech alignment, we employ contrastive learning to maximize the similarity for positive text-speech pairs while minimizing it for negatives. The dot product similarity matrix $B$ is computed as:
$B=\mu{(\mathcal{M}_s \cdot \mathcal{M}_t^{\top})}$, where $\mu$ is a scaling factor. The contrastive loss is then defined as: ${\mathcal{L}_{clap}} = -\log \frac{\exp(B)}{\sum_k \exp(B)}.$

\textbf{Text-Speech Emotion Feature Preparation:} After contrastive pre-training, the frozen EmoCLAP Text and Speech Encoders extract fixed emotion embeddings from text and speech (Fig.~\ref{workflow}, Block 1.2). These embeddings form the retrieval database, supporting both text-speech cross-modal emotion retrieval and emotion representation during speech synthesis.

\subsection{Speaker-Emotion Disentanglement Pre-training}
To disentangle the emotional expression from the speaker's identity and thus serve the subsequent emotion post-correction., we employ adversarial training with a gradient reversal layer (GRL) as the backbone (as shown in Fig.~\ref{workflow} (Block 2)). The EmoCLAP Speech Encoder extracts an emotion representation $\mathcal{H}_e$, which is classified into one of five emotional states using a softmax-based classifier: $P_{\mathcal{H}_e} = C_e(\mathcal{H}_e)$, with the corresponding cross-entropy loss: $\mathcal{L}_e = \mathrm{CE}(P_{\mathcal{H}_e}, \text{emotion\_id}).$

Since $\mathcal{H}_e$ may also encode speaker-specific information, we pass it through a GRL before feeding it into a speaker classifier $C_s$. This yields:
$P_{\mathcal{H}_e}^{s} = C_s(\mathrm{GRL}(\mathcal{H}_e))$.
The adversarial optimization is enforced through an additional cross-entropy loss: $\mathcal{L}_s^{(e)} = \mathrm{CE}(P_{\mathcal{H}_e}^{s}, \text{speaker\_id}),$ ensuring that $ \mathcal{H}_e $ retains only emotion-related information while removing speaker-dependent attributes.

In parallel, the Speaker Encoder extracts a speaker representation $\mathcal{H}_s$, which is used for speaker classification with loss $\mathcal{L}_s$. To ensure that $\mathcal{H}_s$ does not capture emotion-related cues, it is also processed through a GRL prior to an emotion classifier, incurring an additional adversarial loss $\mathcal{L}_e^{(s)}$.
The overall loss function is the sum of these losses: $
\mathcal{L} = \mathcal{L}_s + \mathcal{L}_e + \mathcal{L}_s^{(e)} + \mathcal{L}_e^{(s)}.$

\begin{table*}
  \caption{Emotional comparison analysis before and after the emotion post-correction of the TSE models.}
  \vspace{-1em}
  \label{table1}
  \centering
  \begin{tabular}{p{2.5cm}<{\centering}|p{2cm}<{\centering}p{2cm}<{\centering}|p{1.8cm}<{\centering}p{1.8cm}<{\centering}|p{1.8cm}<{\centering}p{1.8cm}<{\centering}
  }  
    \toprule
        \multirow{2}{*}{\textbf{Method}} & \multicolumn{2}{c|}{\textbf{TSE-MOS}} & \multicolumn{2}{c|}{\textbf{TSEAcc (\%)}} & \multicolumn{2}{c}{\textbf{ECS}} \\
     & \textbf{Before} & \textbf{After} & \textbf{Before} & \textbf{After} & \textbf{Before} & \textbf{After} \\
    \midrule
    Ground Truth                        & NA                         & $4.67 \pm 0.04$              & NA     & 52.1\%          & NA & 1.00 \\
    \midrule
    Editspeech                       & $3.31 \pm 0.03$                         & \textbf{$\mathbf{4.04} \pm \mathbf{0.01}$}     & 8.3\%   & \textbf{47.5\%}   & 0.79 & \textbf{0.97} \\
    $A^3T$                  & $3.48 \pm 0.01$                         & \textbf{$\mathbf{4.13} \pm \mathbf{0.03}$}     & 6.8\%   & \textbf{49.3\%}   & 0.73 & \textbf{0.98} \\
    FluentSpeech                        & $3.17 \pm 0.01$                         & \textbf{$\mathbf{3.93} \pm \mathbf{0.02}$}     & 7.1\%   & \textbf{46.4\%}   & 0.77 & \textbf{0.98} \\
    VoiceCraft              & $3.24 \pm 0.02$                         & \textbf{$\mathbf{4.12} \pm \mathbf{0.02}$}     & 6.9\%   & \textbf{48.8\%}   & 0.71 & \textbf{0.97} \\
    \bottomrule
  \end{tabular}
  \vspace{-1.6em}
\end{table*}

\begin{table}[t]
  \caption{A comparative experiment with different values of K.}
  \vspace{-1em}
  \label{tab4}
  \centering
  \begin{tabular}{p{2cm}<{\centering}p{1.3cm}<{\centering}p{1.3cm}<{\centering}p{1.3cm}<{\centering}} 
    \toprule
    \multirow{2}{*}{\textbf{Method}}  & \multicolumn{3}{c}{\textbf{TSEAcc (\%)}} \\
    & \textbf{K=3} & \textbf{K=5} & \textbf{K=10} \\
    \hline
    Ground Truth                        & 52.1\%                         & 52.1\%                       & 52.1\%                       \\
    \hline
    Editspeech         & \textbf{48.3\% }       & 47.5\%    & 42.5\%                 \\
    $A^3T$               & 47.5\%         & \textbf{49.3\%}     & 43.6\%                 \\
    FluentSpeech         & 46.2\%        & \textbf{46.4\%}       & 41.2\%                  \\
    VoiceCraft          & 48.7\%           & \textbf{48.8\%}    & 44.3\%                 \\
    \bottomrule
  \end{tabular}
  \vspace{-1em}
\end{table}

\begin{table}[t]
  \caption{Comparison of similarity for overall speech quality before and after correction.}
  \vspace{-1em}
  \label{tab5}
  \centering
  \begin{tabular}{p{2.8cm}<{\centering}p{1.8cm}<{\centering}p{1.8cm}<{\centering}}
    \toprule
    \textbf{Method} & \textbf{Energy} & \textbf{MFCC} \\
    \hline
    Editspeech         & 0.91        & 0.96       \\
    $A^3T$             & 0.91        & 0.94     \\
    FluentSpeech       & 0.89        & 0.94      \\
    VoiceCraft         & 0.93        & 0.94    \\
    \bottomrule
  \end{tabular}
  \vspace{-2em}
\end{table}


\subsection{Emotion Post-Correction for TSE}
As illustrated in Fig.~\ref{workflow} (Block 3), the purpose of emotional post-correction is to perform emotional correction on edited speech with inconsistent emotional expressions based on the emotion of the edited text, ensuring that the final speech aligns with the emotional expression of the edited text.

Given an edited text $X_t$, the EmoCLAP Text Encoder first extracts the text emotion embedding $\mathcal{T}_{e}$.
This embedding is then used as a query vector in the \textit{cross-modal emotion retrieval} module, computed by cosine similarity \cite{li2022cross} to retrieve the Top $K$ speech emotion embeddings $ S_K = \{E_{i_1}, E_{i_2}, ..., E_{i_K}\} $ most similar to it.
Furthermore, $ S_K $ are aggregated by averaging them and processed by a pre-trained Adaptor from Block 2 to generate the refined speech emotion embedding $\mathcal{H}_e$.

Simultaneously, the edited speech $X_s$ is processed by the Speaker Encoder to yield the speaker embedding $\mathcal{H}_s$. After that, the $\mathcal{H}_e$ and $\mathcal{H}_s$ are directly added to form a joint embedding, which is then injected to the \textit{Content Adaptor}, \textit{Mel Decoder} and \textit{Flow Post-Net} to integrate emotional characteristics into the synthesized speech. The \textit{Vocoder} synthesizes the final waveform for the given edited text. More details about the TTS synthesizer are referred to \cite{huang2022generspeech}.
The loss function of Block 3 is also consistent with \cite{huang2022generspeech}.



\section{Experiments and Results}


\subsection{Experimental Setup}
We evaluate EmoCorrector on the ECD-TSE.
Precise forced alignment was performed using the Montreal Forced Aligner (MFA) \cite{mcauliffe2017montreal}. The dataset was randomly partitioned into training, validation, and test sets in the proportions of 98\%, 1\%, and 1\%, respectively.
The EmoCLAP Text Encoder extracts a 768-dimensional emotion embedding, while the Speech Encoder extracts an utterance-level 1024-dimensional emotion embedding. These two embeddings are projected into a multimodal space using two learnable projection matrices, resulting in an output dimension of 1024.
The configuration of the Speaker Encoder follows the reference encoder from GST \cite{wang2018style}, comprising a convolutional stack and a GRU block. Each convolutional layer employs a $3 \times 3$ kernel with a stride of $2 \times 2$, followed by batch normalization and ReLU activation. The output channels for the six convolutional layers are $32, 32, 64, 64, 128,$ and $128$, respectively. All classifiers in this architecture consist of a fully connected (linear) layer followed by a softmax layer, and are trained using cross-entropy loss. The pre-trained HiFiGAN \cite{kong2020hifi} vocoder is used to synthesize the speech waveform. An Adam optimizer is used with a learning rate of $1 \times 10^{-7}$.
We set $K$ to 5.
We trained EmoCLAP for 200,000 steps with $\mu$ set to 10 steps, followed by training the speaker-emotion disentanglement model for 300,000 steps. All training was conducted on a single A800 GPU.



\subsection{Evaluation Metrics}
\textbf{Subjective Metric:} \textbf{Text-Speech Emotion Matching Mean Opinion Score (TSE-MOS)} allows listeners to determine whether the emotion of the speech matches the emotion conveyed by the corresponding text.
\textbf{Objective Metrics:} 1) \textbf{Text-Speech Emotion Matching Accuracy (TSEAcc)} is a new LLM-assisted metric. We utilize an advanced Speech Understanding LLM, Qwen2-Audio \cite{Qwen2-Audio} as the speech emotion recognizer to recognize the emotion in speech and then compare it to the emotion label contained in the text to calculate the accuracy. Due to Qwen2's output characteristics, we calculate accuracy group-wise and then average the results.
2) \textbf{Emotional Cosine Similarity (ECS)} adopts emotion2vec \cite{DBLP:conf/acl/MaZYLGZ024} to extract the emotional features of the audio before and after correction and compute the cosine similarity \cite{li2022cross} between these features and the Ground Truth emotional features. The similarity score is in the range of [-1, 1], where a larger value indicates a higher similarity of input samples.

\subsection{Baselines}
We conduct a comparative study of the speech before and after emotion correction using four advanced TSE models, including: 1) \textbf{EditSpeech} \cite{tan2021editspeech} generates text based on the surrounding context to maintain naturalness and quality; 2) \textbf{A}$^3$\textbf{T} \cite{bai20223} proposes alignment-aware acoustic-text pre-training that takes both phonemes and partially-masked spectrograms as inputs; 3) \textbf{FluentSpeech} \cite{jiang2023fluentspeech} uses a diffusion model as the backbone and predicts the masked feature with the help of context speech; and 4) \textbf{VoiceCraft} \cite{DBLP:conf/acl/Peng00MH24} introduces a regression Transformer decoder architecture for neural codec token filling.
We also include \textbf{Ground Truth} speech for comparison.

\subsection{Main Results}
All the pre-training is trained to the convergent state.
In addition, without pre-training, our internal experiments have proved that emotion correction cannot achieve satisfactory results. Due to space constraints, the results of this part of the internal experiments are not displayed. Below we will mainly experiment on the effect of emotion correction and report the results.

In the TSE-MOS evaluation, 25 listeners rated 50 audio samples for the consistency between text and speech emotions. As shown in Table \ref{table1}, EmoCorrector significantly improved the emotional content of the model, with the TSE-MOS score increasing by an average of 0.73 points. Emotion correction for EditSpeech, $A^3T$, FluentSpeech, and VoiceCraft resulted in increases of 0.73, 0.65, 0.76, and 0.88 points, respectively.
In the TSEAcc evaluation, EmoCorrector also enhanced the accuracy of text and speech emotion matching, with an average improvement from 7.2\% to 48\%.
In the ECS evaluation, EmoCorrector made the emotion-corrected speech more similar to the Ground Truth. The emotion correction for EditSpeech, $A^3T$, FluentSpeech, and VoiceCraft improved by 0.18, 0.25, 0.21, and 0.26, respectively.
These results demonstrate that EmoCorrector effectively addresses the issue of emotional inconsistency between edited speech and text.
\subsection{Further Analysis}
\textbf{Analysis of parameter $K$}:
In cross-modal emotion retrieval, the number of retrieved Top $K$ samples may have an impact on the final emotion expression effect.
We set $K$ to \{3,5,10\} to analyze the speech generation results, since 1 may result in insufficient information and a selection between 5 and 10 might not demonstrate a noticeable variance from 5. All the remaining experimental configurations remain the same as before.
As shown in Table \ref{tab4}, the TSEAcc metric achieved the best results with K=5, while performing poorly with 3 and 10. Among these, K=10 yielded the worst outcome, possibly due to the risk of information redundancy when retrieving too many samples.

\textbf{Analysis of the impact of overall quality}:
Although our EmoCorrector focuses on emotion correction, it is not intended to compromise the speech quality.
To validate this, we randomly select 50 test samples. For each sample, we calculate the Energy and MFCC features before and after correction and then obtain their cosine similarity. The results in Table \ref{tab5} show that the similarity values of Energy and MFCC remain around 1, indicating that the model maintains the overall speech quality after emotional correction.




\textbf{Comparison of EVC systems}:
This section validates whether our EmoCorrector has an advantage over using the EVC model directly. We randomly select 50 test samples and adopt two advanced EVC models, that are Cycle-GAN \cite{zhou2020transforming} and VAW-GAN \cite{zhou20d_interspeech}, to conduct emotion conversion for the edited speech. We compare the EVC-converted samples with the emotion-corrected samples, and the experimental results in terms of TSEAcc are shown in Table 6 (Due to limited space, we posted the results on the demo website.). It can be seen from the results that our post-processing method can obtain more accurate synthetic speech of emotions than the EVC method, which has an advantage in emotional rendering.


\vspace{-0.6em}
\section{Conclusion}
This paper presents a novel emotion post-correction scheme for the TSE task, introducing the new benchmarking ECD-TSE dataset and the EmoCorrector that consists of a three-stage pipeline: Text-Speech Emotion Retrieve Database Construction, Speaker-Emotion Disentanglement Pre-training and Emotion Post-Correction for TSE.
Experimental results demonstrate that the proposed framework improves emotional consistency in the edited speech. To the best of our knowledge, EmoCorrector and ECD-TSE are the first methods specifically designed for this task. 
We hope this work provides a foundation for future research in emotion correction for text-based speech editing.

\section{Acknowledgement}
This work was funded by the Young Scientists Fund (No. 62206136), the General Program (No. 62476146) of the National Natural Science Foundation of China, and the Young Elite Scientists Sponsorship Program by CAST (2024QNRC001).
The work by Haizhou Li was supported by the Shenzhen Science and Technology Program (Shenzhen Key Laboratory, Grant No.~ZDSYS20230626091302006), the Shenzhen Science and Technology Research Fund (Fundamental Research Key Project, Grant No.~JCYJ20220818103001002), and the Program for Guangdong Introducing Innovative and Enterpreneurial Teams, Grant No.~2023ZT10X044.


\normalem
\bibliographystyle{IEEEtran}
\bibliography{mybib}

\end{document}